\documentstyle[psfig]{l-aa}
\newcommand{\gm}{$\gamma$-ray }
\newcommand{\egr}{EGRET }
\begin{document}
\thesaurus{11(11.01.2; 11.14.1; 11.17.3); 13(13.07.2)
	}
\title{EGRET gamma-ray source 2EG J0809+5117, a quasar with redshift of 1.14?}
\author{Xue-Bing Wu$^{1,2}$, Qi-Bin Li$^{2}$, Yong-Heng Zhao$^2$ and Li Cao$^2$}
 \institute{1. Institute of Theoretical Physics, Chinese Academy of Sciences, Beijing
 100080, China\\
2. Beijing Astronomical Observatory, Chinese Academy of Sciences, Beijing
 100080, China}
\date{Received  \hspace{2cm}; accepted }
\offprints{Xue-Bing Wu (wuxb@itp.ac.cn)}  
\maketitle
\markboth{X.-B. Wu et al.: EGRET source 2EG J0809+5117, a quasar with redshift 
of 1.14?}{}
   \begin{abstract}  
The low dispersion (400$\AA$/mm) spectrum of the optical counterpart of a flat-spectrum 
radio source 87GB 080315.5+512613, which is one of two possible radio counterparts of 2EG 
J0809+5117, was obtained recently. The optical counterpart, which is $2.02''$ away from
87GB 080315.5+512613 and $19.3'$ away from 2EG J0809+5117, was identified as a quasar with 
redshift of 1.14. We noted that Mattox et al. (1997) suggested the other radio counterpart 
87GB 080459.4+495915 (OJ 508), which is $87.1'$ away from 2EG J0809+5117, is the more potential 
identification, though it was previously suggested to be the 
 identification (with low confidence) of another nearby EGRET source 2EG J0807+4849. Our observation
suggests that it is quite possible that 87GB 080315.5+512613 is the identification of 2EG J0809+5117 rather than 87GB 080459.4+495915. But we still can not exclude the possibility of 87GB 080459.4+495915 at present. Moreover, in order to 
determine whether or not 87GB 080315.5+512613 is a blazar type quasar, the optical 
polarization and variability measures of its optical counterpart are strongly 
encouraged.
\end{abstract}

\section{Introduction}

The Energetic Gamma Ray Experiment Telescope (EGRET) is the high-energy \gm telescope 
on the {\it Compton Gamma-Ray Observatory (CGRO)}. The telescope covers the energy range
 from about 30 MeV to over 20 GeV. From April 1991 to October 1994, the all-sky survey program of EGRET has completed 
 three phases observations.  Up to now, the published
  EGRET catalogs include 157 \gm sources (Fichtel et al. 1994; Thompson et al. 1995; Thompson et al. 1996). Among them, 61 sources have been identified, 
  including 43 AGN with high confidence, 11 AGN with lower confidence, 5 pulsars, one 
  solar flare and Large Magellanic Cloud (LMC). Other 96 sources still remain 
  unidentified.

Except LMC, all previously identified EGRET sources with higher galactic latitude (e.g., 
$|b| >10^o$) are blazar type AGN. These AGN usually have strong, compact, 
flat-spectrum ($\alpha \geq -0.5$, where $S(\nu)\propto\nu^{\alpha}$) radio emission, 
strong optical polarization and significant optical variations on short time scales. 
The blazar class includes objects classified as BL Lacertae type objects, high polarization 
quasars (HPQ), and optically violently variable (OVV) quasars. Although the \gm  radiation 
of blazars has not been well understood, some observational properties of blazars are believed 
to result from a relativistic jet which is directed within $\sim 10^o$ of the line of sight. 

By introducing Bayes' theorem to assess the reliability of the identification of
 EGRET sources with extragalactic radio sources, Mattox et al. (1997) recently 
 demonstrated conclusively that EGRET is detecting the blazar class of AGN. They
  also indicated possible radio identifications of sources with $\mid b \mid >3^o$ 
  in the second EGRET catalog and its supplement. Most of these 
  radio sources have 5GHz radio flux larger than 50 mJy and spectra index larger than -0.5. 
  In order to assure some of these radio sources are more probably the identifications 
  of EGRET sources, optical identifications of these radio sources are necessary. Based 
  on this idea, we are planning a program at Beijing Astronomical Observatory to do the optical 
  spectroscopic studies of the possible optical counterparts of these flat-spectrum 
  radio sources.

\section{2EG J0809+5117 and its radio counterparts}
\begin{center}
\begin{table*}
      \caption{Two possible radio counterparts of 2EG J0809+5117}
         \label{Table 1}
      \[
\begin{array}{cccccccc}
            \hline
            \noalign{\smallskip}
   Name & R.A. & Decl. & S_{5G} & \alpha & r & Contour(\%) & LR\\
           \noalign{\smallskip}
            \hline
           \noalign{\smallskip}
  87GB 080315.5+512613 & 08h07m01.01457s & 51^o17'38.6721'' & 237 & 0.3 & 19.3' & 32.8 & 26.8\\[2mm] \hline
87GB 080459.4+495915 & 08h08m39.66670s & 49^o50'36.5280'' & 1229 & 0.3 & 87.1' & 98.0 & 4.0\\
            \noalign{\smallskip}
            \hline
         \end{array}
      \]
   \end{table*}
\end{center}

\noindent
The main properties of \gm source 2EG J0809+5117 were summarized in the second
EGRET catalog (Thompson et al. 1995). The source position is R.A.=$122.27^o$, Decl.=
$51.29^o$ (J2000) and the Galactic coordinates is $l=167.46^o, b=32.74^o$. Its 
semiminor axes of an eclipse fited to the 95\% confidence 
error contour are 
A=$84'$ and B=$51'$ respectively. The flux (E$>$100MeV)  is
F=9.4 and the 1 $\sigma$ statistical uncertainty in the flux is 
$\triangle$F=2.6 (both in unit of $10^{-8}photons cm^{-2}s^{-1}$). Thompson 
et al. (1995) listed 2EG J0809+5117 as an unidentified source in the second 
EGRET catalog.
Mattox et al. (1997) found two flat-spectrum radio sources from 5GHz radio catalog
(Becker, White \& Edwards 1991) as possible counterparts of
2EG J0809+5117. One is 87GB 080315.5+512613 and the other is 87GB 080459.4+495915. Some 
properties of these two sources are summarized in Table 1. The positions of these two sources
 have been measured with VLA and 
their position errors are about 12 milliarcseconds in 
both
right ascension and declination (Patnaik 
 et al. 1992). 
The source coordinates (R.A. and Decl.) in Table 1 are given in epoch 
J2000.0.  $S_{5G}$ and $\alpha$ 
are the 5GHz flux (in mJy) and spectra index. r is the angle between the radio source and 
the EGRET source. The contour(\%) shows the position confidence contour at the radio position. 
LR is the likelihood ratio indicating the strength of the indication for the identification.

The location of these two sources and the error contour map of 2EG J0809+517 are shown in Figure 
1. 87GB 080459.4+495915 has been identified as a HPQ with z=1.43, namely OJ508. But 
87GB 080315.5+512613 has not been optically identified previously.
Although 87GB 080315.5+512613 is within the 50\% error contour of 2EG J0809+5117 and has larger 
likelihood ratio, Mattox et al. (1997) still listed 87GB 080459.4+495915 as the more potential 
identification of 2EG J0809+5117 based on the probability analyses which gave
the posteriori probability for 87GB 080459.4+495915 and 87GB 080315.5+512613
as 0.079 and 0.014 respectively. This is because that the posteriori probability strongly depends on the priori probability which, for 87GB 080459.4+495915 and 87GB 080315.5+512613, were estimated as 0.0213 and 0.0005. This large difference of the priori probability reflects the fact that strong flat-spectrum radio sources are much more likely to be detected by EGRET. However, we also noted that 
\noindent
87GB 080459.4+495915 
has been suggested as the 
identification (with lower confidence)
of another nearby \egr source 2EG J0807+4849 (Thompson et al. 1993; Nolan et al.
1996). Fig. 1 also shows the locations of 
 there possible radio counterparts and the contour of 2EG J0807+4849. Mattox et al. (1997) also 
 listed 87GB 080459.4+495915 as one of these three possible radio 
counterparts. But they thought another OVV quasar 3C 196, namely 
87GB 080959.9+482202 (z=0.87),
 is the more possible identification of 2EG J0807+4849 than 87GB 080459.4+495915. The other radio 
 source, 87GB 
 080305.0+485033, given by Mattox et al. (1997) as also a possible counterpart of 2EG J0807+4849, 
 is
shown within the 50\% contour of 2EG J0807+4849. It was previously identified as a 
galaxy with visual
magnitude of 18.5 (Becker, White \& Edwards 1991). However, 3C 196 and 87GB 080305.0+485033 
have much steep radio spectra, with spectra index of -0.9 and -0.6 respectively. Because we believe that EGRET is detecting only flat spectrum radio sources, we can not state conclusively that they are the more possible identifications of 2EG J0807+4849 than 87GB 080459.4+495915. Although Mattox et al. (1997) obtained much larger posteriori probability for 3C 196 and 87GB 080305.0+485033 than that for 87GB 080459.4+495915, we must note that this is due to the large difference in the priori probability. They assumed that the priori probability
is only dependent on the 5GHz radio flux. This may bring some errors in estimating the priori probability because the flat-spectrum radio sources are more likely to be detected by EGRET. In this letter, we will not make more discussions about the identification of 2EG J0807+4849 but only focus on the identification of 2EG J0809+5117. For latter case, the optical identification of the other possible flat-spectrum
  radio counterpart of 2EG J0809+5117, namely 87GB 080315.5+512613,
is helpful to see whether or not it could be the true identification.
\begin{figure*}
\vspace*{-0.3cm}
\hspace*{-1.5cm}
\psfig{file=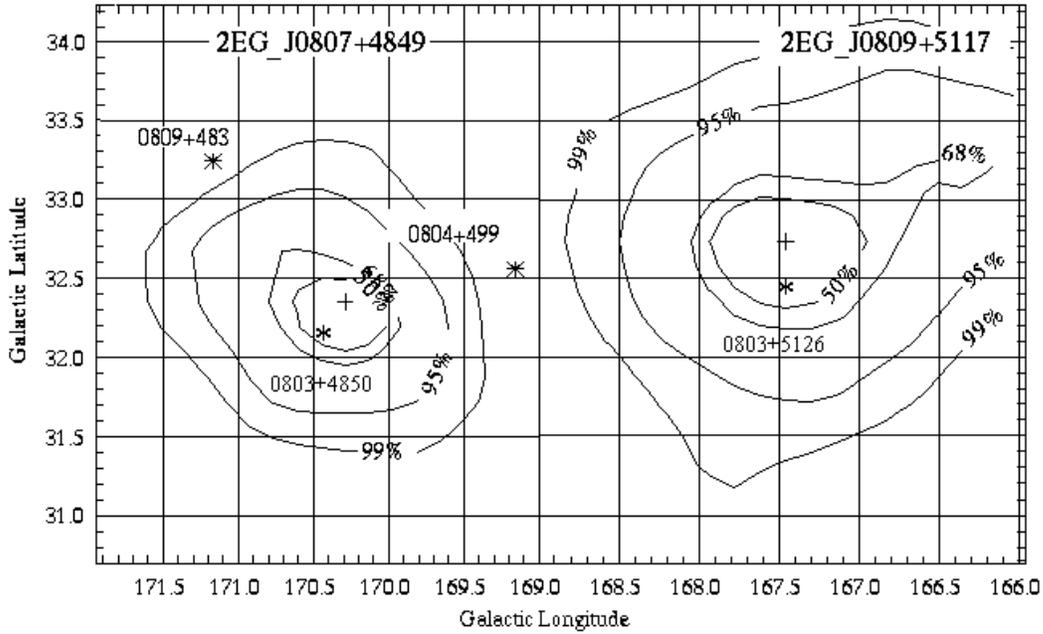,width=17.0cm}
\vspace*{-2.0cm}
\caption[short title]{ The error contours of 2EG J0809+5117 and 2EG J0807+4849 and the locations of 
their possible radio counterparts. The crosses represent the estimated position of EGRET sources 
and the stars represent the position of their possible radio counterparts. 0803+5126, 0804+499, 
0803+4850 and 0809+483 are short for 87GB 080315.5+512613, 87GB 080459.4+495915 (OJ508), 87GB 
080305.0+485033 and 87GB 080959.9+482202 (3C 196) respectively.\\}
\end{figure*}

\section{Probability analysis and spectroscopic observation of the optical counterpart of 
87GB 080315.5+512613}

It is very easy to find the nearest optical source of 87GB 080315.5+512613
 in 
DSS (Palomar Digitized Sky Survey) image. We think this
source, which is only $2.02''$ away from the radio position, is probably the 
optical counterpart of 87GB 080315.5+512613. Its position is R.A.=08h07m01.1s, 
Decl.=$51^o17'40.5''$ (J2000), which is also about $19.3'$ away from the 
estimated position of 2EG J0809+5117. Fig. 2 show its DSS image with size of $10'\times 10'$
and also the 
location of radio position. 
\begin{figure}
\vspace*{-0.4cm}
\hspace*{0.2cm}
\psfig{file=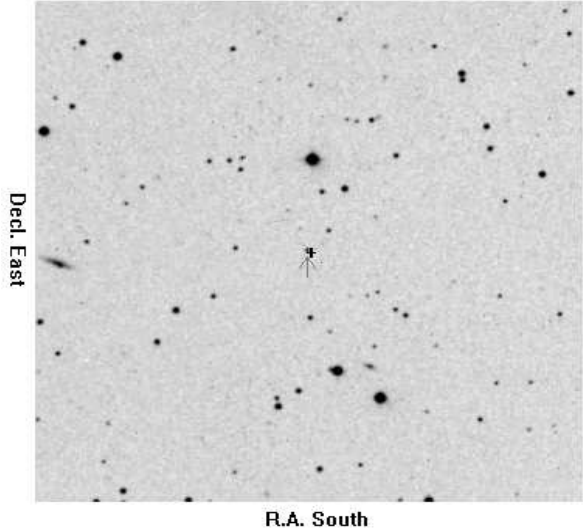,width=8cm,height=7.5cm}
\vspace*{-0.3cm}
\caption[short title]{The DSS image ($10'\times 10'$) centered at the location of 
87GB 080315.5+512613. The arrow points to its optical counterpart and the cross
indicates the radio position.}
\end{figure}

In order to exclude the possibility that this optical source is a confusion source of 87GB 080315.5+512613, we performed a simple quantitative probability analysis. We adopted the method used by de Ruiter, Arp \& Willis (1977) to identify optical counterparts of 1.4GHz radio sources. The posteriori probability of
an optical counterpart can be expressed as $p(id|r)=\frac{\theta}{1-\theta}LR(r)/(\frac{\theta}{1-\theta}LR(r)+1)$ where $\theta$ is the priori probability
and $LR(r)$ is the likelihood ratio given by $LR(r)=\frac{1}{2\lambda}exp(\frac{r^2}{2}(2\lambda-1))$. The dimensionless variables $r$ and $\lambda$ are defined
as $r=(\frac{\Delta\alpha^2}{\sigma_{\alpha}^2}+\frac{\Delta\delta^2}{\sigma_{\delta}^2})^{1/2}$ and $\lambda=\pi\sigma_{\alpha}\sigma_{\delta}\rho(b)$, where $\rho(b)$ is the number density of optical sources, $\Delta\alpha$ and $\Delta\delta$ are the measured position difference between the radio source and an optical object, and
$\sigma_{\alpha}$, $\sigma_{\delta}$ are given by $\sigma_{\alpha}^2=
\sigma_{\alpha_{rad}}^2+{\sigma_o}^2$ and $\sigma_{\delta}^2=
\sigma_{\delta_{rad}}^2+{\sigma_o}^2$. $\sigma_{\alpha_{rad}}$ and 
$\sigma_{\delta_{rad}}$ are the standard deviations of the right ascension and declination positions of the radio source and $\sigma_o$ is the measurement error of the optical position. Patnaik et al. (1992) estimated that $\sigma_{\alpha_{rad}}$ and 
$\sigma_{\delta_{rad}}$ for their VLA measurement are both about 12 milliarcseconds. For optical sources in Palomar Sky Survey, we take a conservative value of $\sigma_o$ as 1 arcsecond. The number density$\rho(b)$ of optical sources in POSS
Plate at galactic latitute of $32.5^o$ is estimated about $6\times 10^{-4}arcsec^{-2}$ (de Ruiter et al. 1977). The position differences $\Delta\alpha$ and $\Delta\delta$, in the sense radio-oprical, are -0.0854 arcsec  and -1.8279 arcsec for our optical counterpart of 87GB 080315.5+512613. Therefore, we obtain that $LR(r)=50.04$. If we assume the priori probability $\theta$ of finding an optical counterpart to a radio source is about 25\%, the posteriori probability for our optical counterpart will be 0.943. Even when we assume $\theta$ is 10\%,
the probability is still as high as 0.917. Therefore, we think that the nearest optical
source to 87GB 080315.5+512613, is most probably its optical counterpart. The
probability that it is a confusion source is very low.

The spectroscopic observations of this source were performed twice on Januray 
5 and 
March 8, 1997 
using 
the CCD detector TEK1024 mounted on the 2.16m reflector at Xinglong station of 
Beijing Astronomical Observatory. The spectra dispersion is about 400$\AA/mm$ 
and
the exposures are 40 and 70 minutes respectively. The spectrum, after the standard light sky 
subtraction 
and the absolute
 flux calibration using MIDAS (Munich Image Data Analysis System developed by ESO),
  is shown in Fig. 3.
\begin{figure}
\vspace*{-0.6cm}
\hspace*{-1.2cm}
\psfig{file=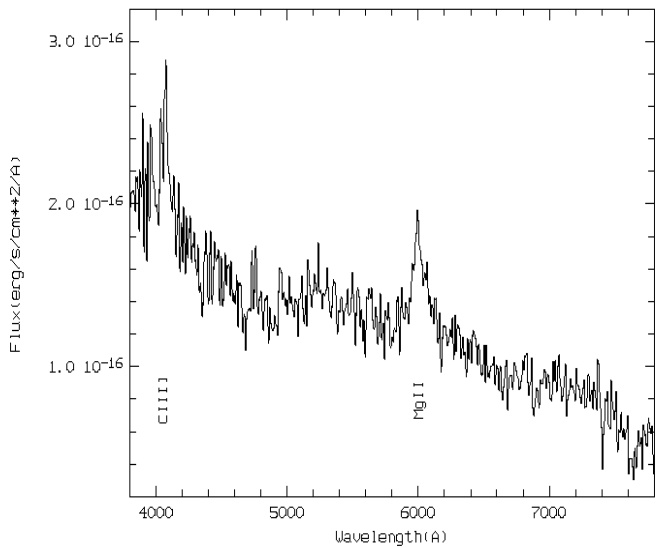,width=10.2cm,height=9.4cm}
\vspace*{-1.8cm}
\caption[short title]{The low dispersion spectrum of the optical counterpart
of 87GB 080315.5+512613.}
\end{figure}

It is clear that the MgII 
 line 
 (2798$\AA$ at rest frame) has shifted to 6008$\AA$ and the
CIII] line (1905$\AA$ at rest frame) shifted to 4080$\AA$. The average redshift
 is 1.14. The 
V-band magnitude, measured at 5500$\AA$, is about 18.5. The luminosity distance
$D_L$ is about 10738$Mpc$ and the V-band absolute magnitude $M_v$ is about
-26.7 (we adopted $H_0=50kms^{-1}Mpc^{-1}$ and $q_0=0$). This quasar has not
been listed in any published catalogs of AGN including a well-known one edited by 
V\'eron-Cetty \& V\'eron (1996).

Because all previously identified EGRET sources with higher galactic latitude are blazars, 
we expect that the optical
source we observed is also a blazar-type quasar if it is really the counterpart
of 2EG J0809+5117. However, it needs to be confirmed by more observations such as 
measuring its optical
variability and polarization. Unfortunately, we are unable to do these 
measurement at Beijing Observatory at present. We strongly encourage
other interested astronomers to do these observations.

\section{Discussions}

We have identified an optical source, which is most probably the counterpart of 
a flat-spectrum radio source 87GB 080315.5+512613, as a quasar with redshift of 
1.14.  Although more optical observations are still
needed to be done
to see whether or not it is a blazar-type quasar, we think
it is quite possible that 87GB 080315.5+512613 is the identification of 2EG J0809+5117 rather than another 
nearby flat-spectrum 
source 87GB 080459.4+495915 (OJ508). The latter one has been previously suggested
to be the lower confidence identification of another EGRET source 2EG J0807+4849 
(Thompson et al. 1993; Nolan et al. 1996). But 
Mattox et al. (1997) recently indicated that another steep spectrum OVV quasar
3C 196 is the more potential identification of 2EG J0807+4849 than 87GB 080459.4+495915 based on the probability analyses. However, in their analyses the priori probability is assumed independent on the EGRET exposure and radio spectra
index, which might bring some errors to the estimated posteriori probability.

Our observation still can not exclude the possibility that 87GB 
080459.4+495915 is the counterpart of 2EG J0809+5117. If the future observations on the optical variability and
polarization of the optical counterpart of 87GB 080315.5+512613 confirm it is a
blazar-type quasar, we think it will enhance the possibility that  87GB 080315.5+512613 is the
identification of 2EG J0809+5117. However, even at this stage we can not
state conclusively that it is the true identification because there are still a lot of blazars not detected by EGRET. Therefor, in order to improve the
present status of the EGRET source identifications, much more observations and analyses are still expected to be done.

\begin{acknowledgements}
We are very much grateful to John Mattox for valuable suggestions on the probability analyses and
helpful comments on the EGRET source identifications. We also thank Jinyao Hu, Jianyan Wei, Jinlin Han and Weimin Yuan for many stimulated 
discussions.
 X.B. Wu acknowledges 
the partial
support from  the Postdoctoral Science Foundation of China. The research has made use of
the NASA/IPAC extragalactic database (NED) which is operated by the Jet Propulsion
Laboratory, Caltech, under contract with the National Aeronautics and Space
Administration.
\end{acknowledgements}


\begin{thebibliography}{}

\bibitem{}Becker, R.H., White, R.L., Edwards,A.L., 1991, ApJS, 75, 1
\bibitem{}de Ruiter, H.R., Arp, H.C., Willis, A.G., 1977, A\&AS, 28, 211
\bibitem{}Fichtel, C.E. et al., 1994, ApJS, 94, 551
\bibitem{}Mattox, J.R. et al., 1997, ApJ, 481, 95
\bibitem{}Nolan, P., et al., 1996, ApJ, 459, 100
\bibitem{}Patnaik, A.R., Browne, I.W.A., Wilkinson, P.N., Wrobel, J.M., 1992, MNRAS, 254, 655
\bibitem{}Thompson, D.J., et al., 1993, ApJS, 86, 629
\bibitem{}Thompson, D.J., et al., 1995, ApJS, 101, 259
\bibitem{}Thompson, D.J., et al., 1996, ApJS, 107, 227
\bibitem{}V\'eron-Cetty, M.-P., V\'eron, P., 1996, ESO, Scientific Report, No. 17
\end{thebibliography}
\end{document}